\documentstyle[twoside,epsfig]{adacta}
\begin{document}
\headings{1}{10}\prvastrana=1\poslednastrana=10
\title{THE QUANTUM TOMOGRAPHIC ROULETTE WHEEL}
\author{G. Mauro D'Ariano \footnote{\email{DARIANO@PV.INFN.IT}} 
and Matteo G. A. Paris \footnote{\email{PARIS@PV.INFN.IT}}}{Dipartimento 
di Fisica 'Alessandro Volta' dell' Universit\'a degli studi di Pavia \\ 
Istituto  Nazionale di Fisica della Materia -- Sezione di Pavia \\ 
via A. Bassi 6,  I-27100 Pavia, Italy}
\def\autor{G. M. D'Ariano and M. G. A. Paris}
\def\nazov{Quantum Tomographic Roulette Wheel}
\abstract{Random-phase homodyne tomography of the field intensity is a
concrete example of the Quantum Roulette of Helstrom. In this paper we
give the explicit POM of such measurement and compare it with direct
photodetection and heterodyne detection. Effects of nonunit quantum
efficiency are also considered. Naimark extensions for the roulette
POM are analyzed and its experimental realization is discussed.} 
\kapitola{I. Introduction}
In recent years much attention has been devoted to state reconstruction 
techniques, namely to measurement schemes providing the elements of the 
density matrix in some representation [1-4]. The problem has a fundamental 
interest and also practical applications as, for example, in determining 
coherence properties of the field and in characterizing effective quantum 
interaction Hamiltonians of optical media. Nevertheless, in many practical 
situations the relevant information one aims to gain about the quantum state 
of a light beam regards the field intensity, namely the photon number. 
Actually, when looking at realistic situations, methods for precise 
measurements of photon intensity are especially welcome. Measurements on 
low-excited highly-nonclassical quantum states of radiation represent, 
in fact, a difficult task, due to limitations of currently available 
photodetectors. On one hand, it is very difficult to discriminate single 
photons in any desidered range of intensity. On the other hand, 
in the low photon-number region ($n < 10$) one can discriminate a single
photon, however with low quantum efficiency. Therefore, alternative methods 
to detect intensity need to be considered, whereas their performances and 
their robustness to low quantum efficiency should be compared in the 
very-quantum regime, versus the limitations of customary photodetection. \par
A way to circumvent photodetection problems is that of amplifying the 
the output photocurrent by mixing the state under examination with a highly 
excited coherent state. This is the basis of homodyne and heterodyne
detection schemes, and, in turn homodyne detection is the basis of
tomographic state-reconstruction. In this paper we address the measurements 
of the field intensity by random-phase homodyne tomography and compare it with 
heterodyne detection. \par
Random-phase homodyne tomography has 
been suggested [6] as an effective tool to measure photon intensity,
and after has received some theoretical attention [5]. More recently 
experiments have nicely demonstrated its ability in determining photon 
statistics for highly nonclassical states of radiation [7]. 
Here, we derive the probability operator measure (POM) that describes the 
random-phase homodyning of the intensity, regarding this measurement 
as a concrete example of the quantum  roulette of Helstrom [8]. 
This abstract framework is suitable to discuss different Naimark
extensions, thus suggesting novel experimental realizations. \par
The paper is structured as follows. In Section II we describe the 
measurement of the field intensity by the quantum tomographic roulette. 
We derive its POM and discuss its precision. In Section III the tomographic 
roulette is compared with direct photodetection and heterodyne detection
also in presence of nonunit quantum efficiency. 
In Section IV we analyze the possible Naimark extensions to the 
roulette POM, whereas Section V closes the paper with some concluding remarks.
\kapitola{II. Measuring Intensity by Quantum Roulette}
The concept of {\em Quantum Roulette} has been introduced by Helstrom 
in his book [8] about twenty years ago. A Quantum Roulette is described 
by a POM  of the form
\begin{eqnarray}
\hat \Pi_m = \sum_{k=1}^{M} \:z_k\: \hat E^{(k)}_m \qquad m=1,...,N
\label{nay1}\:,
\end{eqnarray}
where  $z_k \geq 0$, $\sum_{k=1}^{M}z_k=1$) and the $\hat E^{(k)}_m$'s are 
families of orthogonal projectors corresponding to different observables
labelled by $k$, say $\hat O^{(k)}$, in formula 
\begin{eqnarray}
\hat E^{(k)}_m \hat E^{(k)}_n = \delta_{mn} \hat E^{(k)}_n \qquad 
\sum_{m=1}^{N}E^{(k)}_m=\hat 1
\label{nay0}\:.
\end{eqnarray}
Each experimental event corresponds to the measurement of one of the
observables $\hat O^{(k)}$, chosen at random according to the probability   
distribution ${z_k}$. At the time of Helstrom's proposal the Quantum Roulette 
was just a tool to illustrate an example of generalized measurements
that do not correspond to selfadjoint operators. Nowadays, such measurement 
can be now performed in quantum optics labs. Let us consider
the homodyne detection of a nearly single-mode radiation field. 
When the phase $\phi$ of the local oscillator is fixed, the field-quadrature 
$\hat x_{\phi}=\frac{1}{2} (a^{\dag}e^{i\phi}+ae^{-i\phi})$ is detected and 
the measurement is described by the POM 
\begin{eqnarray}
d\hat E^{(\phi )} (x) = |x\rangle_{\phi}{}_{\phi}\langle x| dx
\label{000}\:, 
\end{eqnarray}
$|x\rangle_{\phi}$ denoting eigenstates of $\hat x_{\phi}$. 
The tomographic detection of the field corresponds to 
scan the local oscillator phase over $[0,\pi ]$.
When the phase $\phi$ of the local oscillator is unknown one deals with
a Quantum Roulette, each experimental event being the measurement  
$d\hat E^{(\phi )} (x)$ with random $\phi$.
The experimental outcomes are
distributed over the whole real axis according to the probability 
distribution $p(x)= \hbox{Tr}\{\hat\varrho d\hat\mu(x)\}$ where 
$d\hat\mu(x)$ is the nonorthogonal POM 
\begin{eqnarray}
d\hat\mu(x) = dx \int_{0}^{\pi} \frac{d\phi}{\pi}
|x\rangle_{\phi}{}_{\phi}\langle x | 
\label{qr1}\:.  
\end{eqnarray}
Inserting the number state expansion of $|x\rangle_{\phi}$
\begin{eqnarray}
|x\rangle_{\phi} = \left(\frac{2}{\pi}\right)^{1/4}
e^{-x^2} \sum_{n=0}^{\infty} \frac{H_{n}(\sqrt{2}x)}{2^{n/2}\sqrt{n!}}       
e^{in\phi} |n\rangle
\label{qr2}\:, 
\end{eqnarray}      
in Eq. (\ref{qr1}) we obtain the POM of the roulette 
\begin{eqnarray}
d\hat\mu(x) = dx \sqrt{\frac{2}{\pi}}e^{-2x^2}\sum_{n=0}^{\infty}
\frac{H^2_{n}(\sqrt{2}x)}{2^n n!} |n\rangle\langle n| = 
dx \sqrt{\frac{2}{\pi}} e^{-2x^2} 
\frac{H^2_{a^{\dag} a}(\sqrt{2}x)}{2^{a^{\dag}a} (a^{\dag}a)!}       
\label{qr3}\:, 
\end{eqnarray}      
where $H_n (x)$ denote Hermite polynomials. \\
The POM $d\hat\mu(x)$ is an operator function of the 
number operator only and the outcome $x$ will be an estimate--generally
biased--of the field intensity. We now proceed in deriving an unbiased
estimate. As it was shown by Richter [9] the expectation value of any 
normally ordered product $\langle a^{\dag n}a^m\rangle$ can be obtained 
from tomographic data by averaging the kernel integral
\begin{eqnarray}
{\cal R}[a^{\dag n}a^m] (x,\phi )= e^{i\phi(m-n)}
\frac{H_{n+m}(\sqrt{2}x)}{2^{(n+m)/2}
\left(\begin{array}{c}n+m\\m\end{array}\right)}
\label{qr4}\:,
\end{eqnarray}      
over the probability distribution $p(x,\phi )$.
For the number operator $\hat n$ Eq. (\ref{qr4}) defines the kernel
\begin{eqnarray}
y \equiv {\cal R}[a^{\dag}a] (x)= 2x^2 - \frac{1}{2}
\label{qr5}\:,
\end{eqnarray}      
which is a phase-independent quantity, hence is suitable for estimation
by a Quantum Roulette. The quantity $y$ traces the field intensity by
averaging over the roulette outcomes distribution
\begin{eqnarray}
\bar{y}= \int_{-\infty}^{\infty}\!\!dx\:{\cal R}[a^{\dag}a](x)\: 
p(x) = \langle\hat n\rangle
\label{qr6}\:.
\end{eqnarray}   
Indeed, Eq. (\ref{qr6}) shows that th function $y(x)$ in Eq. (\ref{qr5})
is the unbiased field intensity estimator for Quantum Roulette. 
The single outcomes $y\equiv {\cal R}[a^{\dag}a](x)$ are random numbers
distributed over the interval $[-1/2,\infty]$. It is clear that the 
determination in Eq. (\ref{qr6}) is meaningful only when also a its
statistical deviation is specified. The latter is given by 
\begin{eqnarray}
\overline{\Delta y} = \sqrt{\overline{y^2}-\bar{y}^2}
\label{qr7}\:, 
\end{eqnarray}   
where 
\begin{eqnarray}
\overline{y^2}= \int_{-\infty}^{\infty}\!\!dx\:{\cal R}^2[a^{\dag}a](x)\: 
p(x) 
\label{qr8}\:.
\end{eqnarray}
The explicit expression of the statistical deviation is given by
\begin{eqnarray}
\overline{\Delta y}^2 = \langle\widehat{\Delta n^2}\rangle
+ \frac{1}{2} \left[\langle\widehat{n^2}\rangle + \langle \hat n\rangle
+1 \right]
\label{qr81}\:,
\end{eqnarray}
$\langle\widehat{\Delta n^2}\rangle$ being the intrinsic
photon number fluctuations.  
We can also specify the whole probability distribution $p(y)$. In fact, from 
the Radon-Nikodym derivative of the roulette POM in Eq. (\ref{qr3}) we
arrive at the roulette POM for field intensity
\begin{eqnarray}
d\hat\mu(y)= \frac{dy}{\sqrt{\pi}} 
\frac{e^{-(y+1/2)}}{\sqrt{y+1/2}}\frac{H^2_{a^{\dag}a}(\sqrt{y+1/2})}
{2^{a^{\dag}a} (a^{\dag}a)!} 
\label{qr9}\:. 
\end{eqnarray}   
\kapitola{III. Quantum Roulette versus Photodetection and Heterodyning}
As it emerges from Eq. (\ref{qr81}) the Quantum Roulette measurement of the
field intensity is noisy, as compared with ideal photodetection. Here
we analyze its performances in the realistic case of nonunit
quantum efficiency. When dealing with $\eta <1 $ the overall output 
noise $\langle\widehat{\Delta n^2}\rangle_{\eta}$ is larger than the 
intrinsic quantum fluctuations $\langle\widehat{\Delta n^2}\rangle$ which 
represents the minimum attainable noise in a measurement of the intensity. 
For non unit quantum efficiency, the roulette POM becomes a Gaussian
convolution of the POM (\ref{qr3})
\begin{eqnarray}
d\hat\mu_{\eta} (y) = \int_{-\infty}^{\infty}\!\!
\frac{dy'}{\sqrt{2\pi}\sigma_{\eta}^2}\: d\hat\mu (y')
\exp\left\{-\frac{(y-y')^2}{2\sigma_{\eta}^2}\right\}
\quad \sigma_{\eta}^2 = \frac{1-\eta}{4\eta}\label{cp1}\:, 
\end{eqnarray}
whereas the unbiased estimator is now given by [10]
\begin{equation}
y_{\eta}\equiv {\cal R}_{\eta} [a^{\dag}a](x)=2x^2-\frac{1}{2\eta}
\label{cp2}\:.
\end{equation}
Inserting Eqs. (\ref{cp1}) and (\ref{cp2}) in Eq. (\ref{qr7}) one has
\begin{eqnarray}
\overline{\Delta y_{\eta}}^2 = \langle\widehat{\Delta n^2}\rangle
+ \frac{1}{2}\langle\widehat{n^2}\rangle + \langle\hat n\rangle \left(\frac{2}{\eta}
-\frac{3}{2}\right) + \frac{1}{2\eta^2} 
\label{cp3}\:.
\end{eqnarray}
This noise has to be compared with the rms variance of direct detection
for nonunit quantum efficiency, which is given by
\begin{eqnarray}
\langle\widehat{\Delta n^2}\rangle_{\eta} =
\langle\widehat{\Delta n^2}\rangle  +
\langle\hat n\rangle \left(\frac{1}{\eta}-1\right)
\label{cp31}\:.
\end{eqnarray}
The difference between $\overline{\Delta y_{\eta}}^2$ and
$\langle\widehat{\Delta n^2}\rangle_{\eta}$ defines the noise
$N_R[\hat n]$ added by Quantum Roulette with respect to direct detection
\begin{eqnarray}
N_R[\hat n] = \frac{1}{2}\left[ \langle\widehat{n^2}\rangle+
\langle\hat n\rangle\left(\frac{2}{\eta}-1 \right) + \frac{1}{\eta^2}\right]
\label{cp32}\:.
\end{eqnarray}
$N_R[\hat n]$ is always positive, and roulette determination is more
noisy than direct photodetection even for nonunit quantum efficiency. \\
Let us now consider heterodyne detection, namely the joint measurement
of two commuting photocurrents, which, in turn, trace a pair of conjugated
field quadratures. Each experimental event corresponds to a point in the
complex plane of the field amplitude and these outcomes are distributed
according to the generalized Wigner function $W_{s} (\alpha,\bar\alpha)$
with ordering parameter $s$ related to the quantum efficiency as
$s=1-2\eta^{-1}$. Starting from the relation
\begin{eqnarray}
\overline{|\alpha|^2} = \int_{\bf C} d^2\alpha \: \alpha\alpha^{*} \:
W_{s} (\alpha,\bar\alpha) =\langle a^{\dag} a \rangle + \frac{1}{\eta}
\label{cp4}\:,
\end{eqnarray}
we are led to consider the shifted square modulus
$I_{\eta}=|\alpha|^2-1/\eta$ as the heterodyne unbiased estimator for the 
field intensity. For unit quantum efficiency $\eta=1$ heterodyne detection 
measures the Husimi $Q$-function $\langle\alpha |\hat\varrho
|\alpha\rangle$, and thus the POM for the field intensity $d\hat\mu (I)$ 
is the marginal one of the Arthurs-Kelly coherent-state POM 
$d\hat\mu (\alpha )=\pi^{-1}|\alpha\rangle\langle\alpha |$, we have
\begin{eqnarray}
d\hat\mu (I) = dI\:\sum_{k=0}^{\infty}\:e^{-(I+1)}\:\frac{(I+1)^k}{k!}\:
|k\rangle\langle k|=  dI \:e^{-(I+1)}\:\frac{(I+1)^{a^{\dag}a}}{(a^{\dag}a)!}
\label{cp5}\:.
\end{eqnarray}
For the determination $\overline{I_{\eta}}=\langle\hat n\rangle$
we need to specify the statistical deviation
\begin{eqnarray}
\overline{\Delta I_{\eta}}^2 = \overline{I_{\eta}^2}-\bar{I_{\eta}}^2
\label{cp6}\:.
\end{eqnarray}   
By using Eq. (\ref{cp4}) and the following relation
\begin{eqnarray}
\overline{|\alpha|^4} = \int_{\bf C} d^2\alpha \: \alpha^2\alpha^{*2} \:
W_{s} (\alpha,\bar\alpha) =  \langle \widehat{n^2} \rangle +
\left(\frac{2}{\eta}-1\right)\langle \hat n \rangle + \frac{1}{\eta^2}
\label{cp7}\:,
\end{eqnarray}
we arrive at the result
\begin{eqnarray}
\overline{\Delta I_{\eta}}^2 =  \langle \widehat{\Delta n^2}\rangle +
\left(\frac{2}{\eta}-1\right) \langle\hat n\rangle+ \frac{1}{\eta^2}
\label{cp8}\:,
\end{eqnarray}
which represents the precision of heterodyne detection in measuring the
field intensity.
From Eqs. (\ref{cp31}) and (\ref{cp8}) we obtain the noise added
by heterodyne detection with respect to direct detection
\begin{eqnarray}
N_H[\hat n]=\frac{1}{\eta}\left[\langle\hat n\rangle +\frac{1}{\eta}\right]
\label{cp81}\:.
\end{eqnarray}
$N_H[\hat n]$ is always a positive quantity, thus also heterodyne
detection is more noisy than direct detection for any value of the
quantum efficiency. \\
A direct comparison between Quantum Roulette and heterodyne detection
can be obtained by considering the difference
$\Delta_{RH}[\hat n]=\overline{\Delta y_{\eta}}^2-
\overline{\Delta I_{\eta}}^2$. From Eqs. (\ref{cp3}) and (\ref{cp8})
one has
\begin{eqnarray}
\Delta_{RH}[\hat n] &=&\langle\widehat{\Delta y_{\eta}^2}\rangle -
\langle\widehat{\Delta I_{\eta}^2}\rangle = \frac{1}{2}\left[\langle
\widehat{n^2}\rangle-\langle\hat n\rangle-\frac{1}{\eta^2}\right]
\nonumber \\ &=& \sum_{n=0}^{\infty}  \frac{1}{2}\left[n^2 - n -
\frac{1}{\eta^2}\right]  \varrho_{nn} 
\label{cp9}\:,
\end{eqnarray}                
$\varrho_{nn}$ being the diagonal elements of the signal density matrix.
$\Delta_{RH}[\hat n]$ has no definite sign, when changing the states 
of radiation. Therefore, it is matter of convenience to choose between the 
two kinds of detection scheme, depending on the state under examination. 
For any value of the quantum efficiency $\eta$ the quantity 
$[n^2 - n -\frac{1}{\eta^2}]$ becomes positive for $n$ larger than
the threshold value 
\begin{eqnarray}
\quad n_{T} = \frac{1+\sqrt{1+4/\eta^2}}{2}
\label{cp10}\:.
\end{eqnarray}      
\begin{figure}[bht]
\begin{center}
\hspace{-25pt}\psfig{file=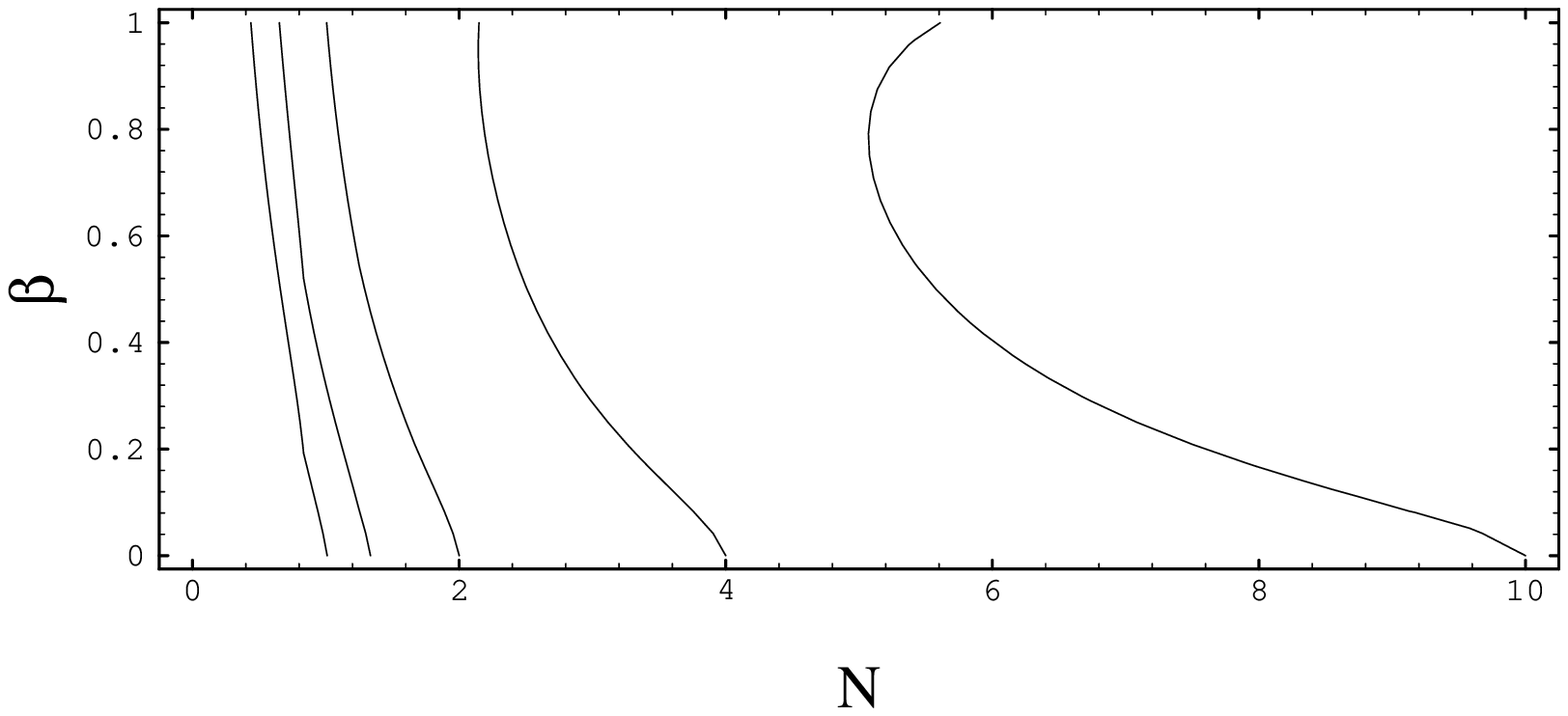,height=7.5cm}
\end{center}
\vspace{-15pt}
\caption{Comparison between the Quantum Roulette and heterodyne detection 
noises in the determination of the field-intensity. The lines where 
$\Delta_{RH}[\hat n]=0 $ for squeezed states are plotted for different
values of the quantum efficiency (from left to right: $\eta =1.0, 0.75,0.5,
0.25\:\hbox{and}\:0.1$) as a function the total mean photon number $N$ and 
the squeezing photon fraction $\beta$. The region on the left of each curve 
corresponds to states for which the Quantum Roulette is more convenient than 
heterodyne detection. This region becomes larger for decreasing $\eta$.}
\label{f:cmp}
\end{figure} 
\par            
This means that the Quantum Roulette is more precise than heterodyne 
detection for low excited states, as the higher-n terms get a lower 
weight $\varrho_{nn}$. Moreover, the lower is the quantum 
efficiency the larger is the threshold value, and hence the region 
where the Quantum Roulette is convenient. This means that the Quantum 
Roulette is very robust to low quantum efficiency.
For coherent states the Quantum Roulette becomes convenient with respect to 
heterodyne detection when the field intensity is lower than the
value $\langle\hat n\rangle = \eta^{-1}$.         
For squeezed states (we consider the case of zero signal and zero squeezing 
phases) Eq. (\ref{cp9}) can be rewritten as 
\begin{eqnarray}
\Delta_{RH}[\hat n] = N^2 + 2\beta N (1+\beta N)+(1-\beta) N (1+2\beta N +2
\sqrt{\beta N (1+\beta N)}) -N -\frac{1}{\eta^2} \nonumber
\label{cp11}\:, 
\end{eqnarray}      
where $N$ denotes the total mean photon number of the state and $\beta$
the squeezing photon fraction, namely the ratio between the (mean) photons 
engaged in squeezing and the total one (for $\beta =0$ we have a 
coherent state and for $\beta =1$ we have squeezed vacuum). 
In Fig. \ref{f:cmp} the lines where $\Delta_{RH}[\hat n]$ is zero are
plotted for different values of the quantum efficiency. 
The region on the left of each curve corresponds to states for which the
Quantum Roulette is more convenient than heterodyne detection.
\kapitola{IV. Naimark Extensions}
Naimark theorem [11] assures that every POM is a partial trace of a 
customary orthogonal projection-valued measure on a larger Hilbert 
${\cal H}={\cal H}_S \otimes {\cal H}_P$ space which, itself, represents 
the original system interacting with appropriate probe mode(s).
This extension is not unique, corresponding to the different possible 
physical implementation of a given measurement. \\ 
For the abstract Quantum Roulette of Helstrom, defined in Eq.
(\ref{nay1}), there exists a standard recipe to obtain a Naimark
extension. This is found by considering the following projectors [8]
\begin{eqnarray}                  
\hat E_m  = \sum_{k=1}^{M}\:\hat E^{(k)}_m  \:\otimes\:
|\omega_k\rangle\langle\omega_k |
\label{nay2}\:,
\end{eqnarray}
where $|\omega_k\rangle$ denotes an orthogonal set of states in the
extension (probe) ${\cal H}_P$ Hilbert space.
In fact, by preparing  the probe in the state
\begin{eqnarray}
|\psi_P\rangle = \sum_{k=1}^{M}\:z_k^{1/2}\:  |\omega_k\rangle
\label{nay3}\:,
\end{eqnarray}
we have
\begin{eqnarray}
\hbox{\rm Tr}_P \left\{ \hat 1_S \otimes |\psi_P\rangle\langle\psi_P|
\: \hat E_m \right\} = \hat \Pi_m
\label{nay4}\:.
\end{eqnarray}
The POM in Eq. (\ref{qr1}) is the continuous version of the quantum
roulette with $\phi$ playing the role of the label $k$.
The Helstrom recipe to achieve a Naimark extension requires an
orthogonal POM for the phase $\phi$ in ${\cal H}_P$ yielding a
resolution of identity in $[-\pi,\pi]$. As it describes a phase
variable, such a POM has to satisfy the additional requirement of 
covariance, namely
\begin{eqnarray}
d\hat\mu (\phi) =\frac{d\phi}{2\pi}\:e^{-i\hat S\phi}\:\hat P\:e^{i\hat S\phi}
\label{nay5}\:,
\end{eqnarray}
$\hat S$ being the phase-shift generator in the probe Hilbert space,
and $\hat P$ a suitable positive operator.  \\
The "minimal" implementation for the roulette POM would be a two-mode
system, with the additional mode playing the role of the probe. However, 
such an implementation cannot be fully quantum, because no single-mode 
orthogonal set the phase variable exists. By using the canonical (Susskind-Glogower) phase POM
$d\hat\mu (\phi)=(2\pi)^{-1}d\phi\sum_{nm}\exp\left[i(n-m)\phi\right]|n\rangle
\langle m|$ we have an approximated extension described by the two-mode
(non-orthogonal) POM
\begin{eqnarray}
d\hat M(x) = dx \int\frac{d\phi}{2\pi}\:|x\rangle_{\phi}{}_{\phi}\langle
x|\:\otimes\:\sum_{nm}\exp\left[i(n-m)\phi\right]\:|n\rangle\langle m|
\label{nay6}\:,
\end{eqnarray}
which corresponds to the measured photocurrent
\begin{eqnarray}
\hat X = \int x \: d\hat M(x) = a^{\dag} \hat e_- + a \hat e_+
\label{nay7}\:,
\end{eqnarray}
$\hat e_{\pm}$ being the raising and lowering operators on ${\cal H}_P$
\begin{eqnarray}
\hat e_+ = b^{\dag}\:\frac{1}{\sqrt{b^{\dag}b+1}}\qquad
\hat e_- = \frac{1}{\sqrt{b^{\dag}b +1}}\:b
\label{nay8}\:.
\end{eqnarray}
An exact two-mode implementation can be achieved in the semiclassical
limit of highly excited probe mode. The basic idea is that coherent states
$|z\rangle$ provide exact phase states for high amplitude
$|z|\rightarrow\infty$. In this limit we can neglect the ordering in
(\ref{nay8}) and writing
\begin{eqnarray}
a^{\dag} \hat e_- + a \hat e_+  \simeq
\frac{a^{\dag} b + ab^{\dag}}{\sqrt{b^{\dag}b +1}} 
\simeq
\frac{a^{\dag} b- + ab^{\dag}}{\sqrt{\langle b^{\dag}b \rangle}}
\label{nay9}\:,
\end{eqnarray}
which coincides with the homodyne photocurrent.
The prescription (\ref{nay3}) for the probe preparation becomes
\begin{eqnarray}
\hat\varrho_P= \lim_{|z|\rightarrow\infty}\:\int\frac{d\phi}{2\pi}\:
\left| |z|e^{i\phi}\rangle\langle |z|e^{i\phi} \right|
\label{nay10}\:.
\end{eqnarray}
Eqs. (\ref{nay9}) and (\ref{nay10}) unambiguously identify random-phase
homodyne detection as the minimal (semiclassical) implementation of the 
Quantum Roulette POM of Eq. (\ref{qr1}). \\
A fully quantum extension requires more than one mode as the probe. 
An example is provided by the heterodyne phase eigenstate, which are
defined on a two-mode Hilbert space [12]. 
\kapitola{V. Conclusions}
The Quantum Roulette is a concrete example of the Quantum Roulette 
of Helstrom. In this paper we compared intensity measurement 
by the Quantum Roulette with that from direct photodetection and from 
heterodyne detection. Direct photodetection, though only in principle,
remains the most precise detection scheme, also in the case of nonunit
quantum efficiency. On the other hand, the choice between Quantum Roulette 
and heterodyne detection is a matter of convenience, depending on the state 
under examination. For coherent states, the Quantum Roulette becomes 
convenient with respect to heterodyne detection when the field intensity 
is lower than threshold value $\langle\hat n\rangle = \eta^{-1}$.
In general, Quantum Roulette is more convenient for low excited states,
and is more robust in presence of low efficient photodetectors.  
\kapitola{Acknowledgements}
The work of M.G.A.P. is supported by a post-doctoral grant of the 
University of Pavia. 
\kapitola{References}
\begin{description}
\item{[1]} \refer{D. T. Smithey, M. Beck, M. G. Raymer, A. Faridani}{Phys. Rev.
 Lett.} {70}{1993}{1244}
\item{[2]} \refer{G. M. D'Ariano, C. Macchiavello, M. G. A. Paris}{Phys.
Rev. A}{50}{1994}{4298} \refer{G. M. D'Ariano, U. Leonhardt, H.
Paul}{Phys. Rev. A}{52}{1995}{1801} \refer{G. M. D'Ariano, C. Macchiavello, 
M. G. A. Paris}{Phys. Lett. A}{195}{1994}{31}
\item{[3]} \refer{K. Banaszek, K. Wodkiewicz}{Phys. Rev .Lett}{76}{1996}{4344} 
\item{[4]} \refer{H. Paul, P. T\"{o}rma, T. Kiss, I. Jex}{Phys. Rev. Lett.}
{76}{1996}{2464}
\item{[5]} \refer{K. Banaszek, K. Wodkiewicz}{Operational Theory of
Homodyne detection}{Atom-Ph/9609001}{1996}{}
\item{[6]} \refer{M. Munroe, D. Boggavarapu, M. E. Anderson, M. G.
Raymer}{Phys. Rev. A}{52}{1995}{R924}        
\item{[7]} \refer{S. Schiller, G. Breitenbach, S. F. Pereira, T. M\"{u}ller,
J. Mlynek}{Phys. Rev. Lett.}{77}{1996}{2933}
\item{[8]} C. W. Helstrom, {\sl Quantum Detection and Estimation Theory} 
(Academic Press, New York) (1976).  
\item{[9]} \refer{Th. Richter}{Phys. Lett. A}{221}{1996}{327}
\item{[10]} \refer{G. M. D'Ariano, M. G. A. Paris}{Added noise in
homodyne measurement of field-observables}{Quant-Ph/9704034}{1997}{}
\item{[11]} \refer{M. A. Naimark}{Izv. Akad. Nauk SSSR Ser.Mat.}{4}{1940}{227};
 W. Mlak, {\it Hilbert Spaces and Operator Theory} (Kluwer Academic, Dordrecht)(1991).                     
\item{[12]} \refer{G. M. D'Ariano, M. F. Sacchi}{Phys. Rev. A}{52}{1995}{R4309}
\end{description}
\end{document}